\documentclass[pre,aps,twocolumn,superscriptaddress]{revtex4-1}
\pdfoutput=1
\usepackage{amssymb}
\usepackage{epsfig}
\usepackage{amsmath}
\usepackage{subfigure}
\usepackage{graphicx}
\usepackage{textcomp}
\usepackage{url}
\usepackage{float}
\usepackage{color}
\usepackage{cases}
\usepackage[normalem]{ulem}

\usepackage[titletoc,title]{appendix}

\usepackage[colorlinks=true, urlcolor=blue, anchorcolor=blue, citecolor=blue,filecolor=blue,linkcolor=blue,menucolor=blue
]{hyperref}

\usepackage[colorlinks=true, urlcolor=blue, anchorcolor=blue, citecolor=blue,filecolor=blue,linkcolor=blue,menucolor=blue
]{hyperref}

\usepackage{color}

\usepackage[titletoc,title]{appendix}

\begin{document}

\title{Thermally activated particle motion in biased correlated Gaussian disorder
potentials}

\author{Alexander Valov}
\email{aleksandr.valov@mail.huji.ac.il}
\affiliation{Racah Institute of Physics, Hebrew University of Jerusalem, Jerusalem 91904, Israel}

\author{Netanel Levi}
\email{netanel.levi2@mail.huji.ac.il}
\affiliation{Racah Institute of Physics, Hebrew University of Jerusalem, Jerusalem 91904, Israel}

\author{Baruch Meerson}
\email{meerson@mail.huji.ac.il}
\affiliation{Racah Institute of Physics, Hebrew University of Jerusalem, Jerusalem 91904, Israel}

\begin{abstract}
Thermally activated particle motion in disorder potentials is controlled
by the large-$\Delta V$ tail of the distribution of height $\Delta V$ of the potential barriers created by the disorder.  We employ the optimal fluctuation method to evaluate
this tail for correlated quenched Gaussian potentials in one
dimension in the presence of a small bias of the potential. We focus on the mean escape time (MET)
of overdamped particles averaged over the disorder.
We show that the bias leads to a strong (exponential) reduction of the MET in the direction along the bias. The reduction
depends both on the bias, and on detailed properties of the covariance
of the disorder, such as its derivatives and asymptotic behavior at large distances.
We verify our theoretical predictions for  the large-$\Delta V$ tail of the barrier height distribution, as well as earlier predictions of this tail for zero bias, by performing large-deviation simulations of the potential disorder. The simulations employ correlated random potential sampling based on the circulant embedding method and the Wang-Landau algorithm, which enable us to probe probability densities smaller than $10^{-1200}$.
\end{abstract}
\maketitle

\section{Introduction}

Slow thermally activated motion of overdamped particles in a
quenched disorder potential is an important research paradigm,
which is relevant in many applications such as supercooled liquids and glassy
matrices \citep{solidstate1,solidstate2,Bassler,glass1}, the motion of
particles in disordered metals or semiconductors \citep{glass2,glass3}
the motion of macromolecules in DNA \citep{bio1,bio2,bio3}, \textit{etc}. Direct experiments
with this system have recently become available in the form of laser-produced quenched random potentials
in colloids \citep{G1,G2,G3,G4,speckle}. Since the pioneering works
of DeGennes \citep{DeGennes} and Zwanzig \citep{Zwanzig}, there have been many theoretical
studies in this direction \citep{LV,anomalous1,anomalous2,anomalous3,anomalous4,Wilkinson,M2022}, to name but a few.

When the thermal noise is small, the mean escape time (MET) of particles from a local potential well of the disordered potential is determined by the large-$\Delta V$ tail of the probability distribution $\mathcal{P}(\Delta V)$ of the \emph{potential barriers} $\Delta V$ created by the disorder. This tail can be efficiently evaluated by using the optimal fluctuation method (OFM) \citep{LV,M2022,aboutLV}. In particular, it was found in Ref. \cite{M2022} that this tail strongly (exponentially) depends on whether the covariance of the disorder decreases monotonically  with the distance or not. These findings, however, were limited to the unbiased case, that is when the ensemble average of the random potential $V(x)$ at any $x$ is zero. In experimental situations there can be a systematic external potential bias, and it is interesting to investigate its role.  Here we show that  the bias can strongly affect the large-$\Delta V$ tail of the distribution $\mathcal{P}(\Delta V)$ and, as a result, lead to an exponentially strong reduction of the MET in the direction along the bias. This reduction depends
both on the bias, and on detailed properties of the covariance of the disorder, such as its
derivatives and asymptotic behavior at large distances. The OFM calculations are based on the determination of the optimal -- that is, the most likely -- configuration of the random potential $V(x)$ which dominates the large-$\Delta V$ tail of $\mathcal{P}(\Delta V)$ \citep{LV,M2022}.

To verify our theoretical predictions for  the large-$\Delta V$ tail of the barrier height distribution, as well
as the earlier predictions of this tail for zero bias ~\cite{LV,M2022}, we perform large-deviation simulations of the potential disorder. The simulations employ correlated random potential sampling based on the circulant
embedding method and the Wang-Landau algorithm, which enable us to probe probability densities
smaller than $10^{-1200}$. As we will show, the simulation results strongly support the theory.

Let us introduce the basic model that we consider in this work.
Overdamped particle motion in a quenched
disorder potential $V\left(x\right)$ can be described by the Langevin equation
\begin{equation}
\dot{x}=-\mu\frac{dV(x)}{dx}+\sqrt{2D}\xi(t)\,,\label{Langevin}
\end{equation}
where $\mu$ is the mobility, $D$ is the diffusion coefficient of the particle in the absence of the potential, and $\xi(t)$
is a delta-correlated Gaussian noise with zero mean. In the following we set $\mu=1$ (which renders somewhat unusual units to the potential, $[V]=\text{length}^2/\text{time}$).

We suppose that the quenched random potential $V\left(x\right)$ is  statistically homogeneous in space and normally
distributed. The  potential barrier $\Delta V$ is formally defined as $\Delta V=V\left(x=L\right)-V\left(x=-L\right)$
where we assume, without limiting the generality, that $x=-L$ is a minimum point of $V(x)$,  $x=L$ is a maximum point, and $V'(x)>0$ for all $|x|<L$. In this case the activated escape proceeds from left to right.

For a fixed realization of the potential $V(x)$, the MET over this barrier, averaged over the realizations of the thermal noise, is described  by the classical Kramer's formula \citep{Kramers}:
\begin{equation}
T\sim\exp\left(\frac{\Delta V}{D}\right)\,,\label{Kramerstime}
\end{equation}
where we have omitted the pre-exponential factor which we will not be interested in.  As in the previous works \cite{LV,M2022}, we will focus on the MET additionally averaged over different realizations of the disorder potential $V(x)$. We will denote it by $\langle T \rangle$.   In the limit of  $D\rightarrow0$, $\langle T \rangle$ is controlled
by the large-$\Delta V$ tail of the barrier distribution $\mathcal{P}\left(\Delta V\right)$ \cite{LV,M2022}. This
tail can be represented as
\begin{equation}
\mathcal{P}\left(\Delta V\rightarrow\infty\right)\sim\exp\left[-S\left(\Delta V\right)\right]\,,\label{tailgeneral}
\end{equation}
where $S\left(\Delta V\right)$ is a large-deviation function that will be in the focus of our attention.  Therefore,
\begin{equation}
\left\langle T\right\rangle \sim\int_{0}^{\infty}\exp\left[\frac{\Delta V}{D}-S\left(\Delta V\right)\right]d\left(\Delta V\right)\,.\label{meanestimeaverage}
\end{equation}
Since $D\to 0$, this integral can be evaluated via the Laplace's method. The saddle point $\Delta V_{*}$ is the maximum point of the function
\begin{equation}
\phi\left(\Delta V\right)=\frac{\Delta V}{D}-S\left(\Delta V\right)\,.\label{phi}
\end{equation}
As a result, the MET averaged over the realizations of disorder can be evaluated as
\begin{equation}
\left\langle T\right\rangle \sim\exp\left[\frac{\Delta V_{*}}{D}-S\left(\Delta V_{*}\right)\right]\,.\label{Tdoubleav1}
\end{equation}
To implement the evaluation, outlined in Eqs.~(\ref{phi}) and (\ref{Tdoubleav1}), we first need to determine the large-deviation function $S\left(\Delta V\right)$. These calculations are presented
in Sec. \ref{OFM}. The simulation algorithm is briefly described in  Sec. \ref{Details}, and Sec. \ref{Results} presents the simulations results. A brief summary, discussion and possible extensions of our results are given in Section \ref{SD}.

\section{\label{OFM} Optimal Fluctuation Method}

A statistically homogeneous random Gaussian potential $V(x)$ is fully determined by its mean (which describes the bias if there is one, see below) and the covariance
\begin{equation}
\kappa\left(x-x'\right)=\left\langle V\left(x\right)V\left(x'\right)\right\rangle -\langle V(x) \rangle \langle V(x') \rangle \,.\label{correlatorgeneral}
\end{equation}
We will assume that $\kappa(z)$ has its absolute maximum at $z=0$ and is at least twice differentiable. $\kappa(z)$ can be either a monotonically decreasing  function of $z$, or nonmonotonic  \cite{difference}. The variance of $V(x)$ is equal to $\kappa\left(0\right)$.

In the absence of bias, the statistical weight of a realization of the Gaussian disorder potential
$V(x)$ is determined by the action functional \citep{Zinn-Justin}
\begin{equation}
\mathcal{S}\left[V\left(x\right)\right]=
\frac{1}{2}\int_{-\infty}^{\infty}dx \int_{-\infty}^{\infty}dx' K\left(x-x'\right)
V\left(x\right)V\left(x'\right)\,,\label{actiongeneral}
\end{equation}
where $K\left(x-x'\right)$ is the inverse kernel, 
defined by the equation
\begin{equation}
\int_{-\infty}^{\infty}dx'' K\left(x-x''\right)
\kappa\left(x'-x''\right)=\delta\left(x-x'\right)\,.\label{inverse}
\end{equation}

In the presence of a uniform bias field, $E=\text{const}$,  we have $\langle V(x)\rangle = -Ex$, and the potential $V(x)$ can be represented as
\begin{equation}
V\left(x\right)=v\left(x\right)-Ex\,,\label{beginning_potential_with_bias}
\end{equation}
where $v\left(x\right)$ is a normally distributed random field with zero mean and
the covariance $\kappa\left(z\right)$.

The large-$\Delta V$ tail of $\mathcal{P}\left(\Delta V\right)$ describes atypically
large barriers,  which are dominated by the optimal configuration
of the potential $V(x)$ conditioned on the specified $\Delta V$ \cite{LV,M2022}. The optimal configuration minimizes, subject to additional conditions that we will specify shortly, the action functional \citep{Zinn-Justin}:
\begin{eqnarray}
\mathcal{S}\left(v,E\right)&=&\frac{1}{2}\int_{-\infty}^{\infty}dx \int_{-\infty}^{\infty}dx'K\left(x-x'\right)v\left(x\right)v\left(x'\right) \nonumber\\
 & =&\frac{1}{2}\int_{-\infty}^{\infty}dx \int_{-\infty}^{\infty}dx'K\left(x-x'\right)
 \nonumber\\
 &&\times\left[V\left(x\right)+Ex\right]\left[V\left(x'\right)
 +Ex'\right]\,.\label{first_action_with_bias}
\end{eqnarray}
Assuming that the optimal configuration of $V\left(x\right)$ is smooth,
we can write down the conditions specifying the potential barrier $\Delta V$ on an interval $|x|<L$ of an a priori unknown length $2L$ \cite{arbitraryconstant}:
\begin{eqnarray}
 &  & V\left(x=L\right)-V\left(x=-L\right)=\Delta V\,,\label{constraints}\\
 &  & \frac{dV}{dx}\left(x=-L\right)=\frac{dV}{dx}\left(x=L\right)=0\,,\label{defminusL}\\
 &  & \frac{d^{2}V}{dx^{2}}\left(x=-L\right)>0,\ \frac{d^{2}V}{dx^{2}}\left(x=L\right)<0\,,\label{defL}\\
 & & \frac{dV}{dx}>0\,,\quad |x|<L\,,\label{monotonV}
\end{eqnarray}
where the inequality (\ref{monotonV}) guarantees that there are no other extrema  of $V\left(x\right)$ on the interval $|x|<L$.

Accommodating the constraint (\ref{constraints}) via a Lagrange multiplier $\lambda$ and two delta-functions, we obtain a modified  action functional to be minimized:
\begin{eqnarray}
s_{\lambda}\left[V \left(x\right)\right]&=\frac{1}{2}\int_{-\infty}^{\infty}dx \bigg\{ \int_{-\infty}^{\infty}dx'K\left(x-x'\right)\nonumber \\
&\times\left[V\left(x\right)+Ex\right]\left[V\left(x'\right)+Ex'\right]\nonumber \\
&-\frac{\lambda}{2} V\left(x\right)\left[\delta\left(x-L\right)-\delta\left(x+L\right)\right]\bigg\}\,.
\label{functional_with_m}
\end{eqnarray}
An explicit account of the constraints (\ref{defminusL})-(\ref{monotonV}) in the action minimization procedure is quite difficult. Therefore, we will proceed without accounting for these constraints, and make sure \textit{a posteriori} that they are obeyed by the solution.

The linear variation of the action functional (\ref{functional_with_m}) must vanish, so that
\begin{eqnarray}
  &&\delta s_{\lambda}\left[V\left(x\right)\right]
  \!=\!\frac{1}{2}\int_{-\infty}^{\infty}dx\, \delta V\left(x\right)\bigg\{ \int_{-\infty}^{\infty}dx' K\left(x-x'\right)\nonumber \\ &&\times \left[V\left(x'\right)+Ex'\right]
 \!-\! \frac{\lambda}{2}\left[\delta\left(x-L\right)-\delta\left(x+L\right)\right]\bigg\}=0, \label{functional_with_m_var}
\end{eqnarray}
leading to the linear integral equation
\begin{eqnarray}\label{inteq1}
\int_{-\infty}^{\infty}K\left(x-x'\right)
\left[V\left(x'\right)+Ex'\right]dx' \nonumber \\
= \frac{\lambda}{2}\left[\delta\left(x-L\right)-\delta\left(x+L\right)\right]\,.
\end{eqnarray}
Comparing this equation with Eq.~(\ref{inverse}), one can easily guess the solution:
\begin{equation}
V\left(x\right)=\frac{\lambda}{2}\left[\kappa\left(x-L\right)
-\kappa\left(x+L\right)\right]-Ex .\label{potential_after_sol_guess}
\end{equation}
Then, using Eqs.~(\ref{constraints}) and~(\ref{defminusL}), we determine the Lagrange multiplier $\lambda$,
\begin{equation}
\lambda=\frac{\Delta V+2EL}{\kappa\left(0\right)-\kappa\left(2L\right)}\,, \label{lambda_after_sol_1}
\end{equation}
and obtain an (in general, transcendental) equation for the optimal value of $L$:
\begin{equation}
\kappa'\left(2L\right)=-\frac{2E\left[\kappa\left(0\right)-\kappa\left(2L\right)\right]}{\Delta V+2EL}\,,\label{close_eq_1}
\end{equation}
where $\kappa'(\dots)$ is the derivative of the covariance with respect to its argument.

Therefore, the optimal, \textit{i.e.} the least  improbable, configuration of the disorder potential $V(x)$, conditioned on the large potential barrier $\Delta V$, is the following:
\begin{equation}
V(x)\!=\!\frac{\Delta V+2EL}{2\left[\kappa\left(0\right)-
\kappa\left(2L\right)\right]}\left[\kappa\left(x-L\right)
\!-\!\kappa\left(x+L\right)\right]\!-\!Ex .\label{potential_after_sol_1}
\end{equation}
We have already imposed the conditions (\ref{constraints}) and~(\ref{defminusL}). Taking the second derivative of the both sides of Eq.~(\ref{potential_after_sol_1}) we see  that, for $E>0$ (which we should require in any case, see below),
the conditions~(\ref{defL}) are also satisfied. The fulfillment of the monotonicity condition~ (\ref{monotonV}) depends
on the  specific form of covariance, and we will discuss it shortly.  Meanwhile, using Eqs.~(\ref{actiongeneral}) and  (\ref{potential_after_sol_1}), we can calculate the action:
\begin{equation}
S\left(\Delta V,L\right)=\frac{\left(\Delta V+2EL\right)^{2}}{4\left[\kappa\left(0\right)-\kappa\left(2L\right)\right]}\,,\label{action_after_sol_1}
\end{equation}
where $L$ is the solution of Eq.~(\ref{close_eq_1}).
In the following subsections we will consider several cases  depending on the form of the covariance and on the sign and magnitude of the external bias.

\subsection{Zero bias}
\label{zerobias}

Let us first briefly review the zero-bias case, $E=0$, previously considered in Ref. \cite{LV,M2022}, and highlight the crucial difference between monotonically decreasing (MD)  and nonmonotonic (NM) covariances, uncovered in Ref. \cite{M2022}. Where necessary, we will also distinguish between nonmonotonic covariances that become negative at some distances -- nonmonotonic negative (NMN) for brevity,  and nonmonotonic but everywhere positive (NMP) covariances. In the absence of bias, Eq.~(\ref{potential_after_sol_1}) for the optimal configuration of the potential gives \cite{M2022}
\begin{equation}
V\left(x\right)=\frac{\Delta V}{2\left[\kappa\left(0\right)-
\kappa\left(2L\right)\right]}\left[\kappa\left(x-L\right)-
\kappa\left(x+L\right)\right]\,,
\label{V2}
\end{equation}
and the action~(\ref{action_after_sol_1}) becomes \cite{M2022}
\begin{equation}
S\left(\Delta V\right)=\frac{\left(\Delta V\right)^{2}}{4\left[\kappa\left(0\right)-\kappa\left(2L\right)\right]}\,.\label{act}
\end{equation}

For the MD covariance $\kappa(z)$ the action (\ref{act}) is a monotonically decreasing function of $L$. As a result, the minimum action is achieved in the limit of $L\to\infty$. That is, the optimal configuration of $V(x)$ in this case has the form of an isolated pair of a \textit{spike} and an \textit{antispike}, whose shape is determined by the shape of the covariance $\kappa(z)$ \cite{LV,M2022}.

For the NM covariance the situation is different. Let us denote by $\ell_*>0$ the closest to zero position of the local \emph{minimum} of $\kappa(z)$.
To minimize the action (\ref{act}) (at least locally) and satisfy the conditions (\ref{defminusL})-(\ref{monotonV}), we can set $L=\ell_*/2$. The optimal configuration of the disorder potential in this case is localized \cite{M2022}.
Under these assumptions, and using Eq.~(\ref{tailgeneral}), we obtain the following predictions for the large-$\Delta V$ tail of the potential barrier distribution $\mathcal{P}(\Delta V\to\infty)$  in the two cases \cite{LV,M2022}:
\begin{equation}
\!-\!\ln\mathcal{P}(\Delta V)\!\simeq\!
    \begin{cases}
         \frac{\left(\Delta V\right)^{2}}{4\kappa\left(0\right)} & \text{for MD covariance,}\\
         \frac{\left(\Delta V\right)^{2}}{4\left[\kappa\left(0\right)-\kappa\left(\ell_*\right)\right]} & \text{for NM covariance.}
    \end{cases}
\label{PdV}
\end{equation}
In its turn, the saddle-point evaluation, outlined in Eqs.~(\ref{tailgeneral})-(\ref{Tdoubleav1}) yields the MET averaged over disorder:
\begin{equation}
\ln\left\langle T\right\rangle \simeq
    \begin{cases}
         \frac{\kappa\left(0\right)}{D^2} & \text{for MD covariance} \\
           \frac{\kappa\left(0\right)-\kappa\left(\ell_*\right)}{D^2} &
           \text{for NM covariance}
    \end{cases}
\label{Taveragemon}
\end{equation}
Because of the very large $1/D^2$ factor in Eqs.~(\ref{Taveragemon}), the MET averaged over the disorder is extremely long \citep{LV,M2022}. A similar  exponential suppression, but observed in the long-time particle diffusion in disordered potentials, has been known for a long time \cite{DeGennes,Zwanzig}.

Another striking effect, described by Eqs.~(\ref{Taveragemon}), is specific to the averaged-over-disorder MET $\langle T \rangle$. It describes a very strong (exponential) dependence of $\langle T \rangle$ on whether the covariance $\kappa(z)$ is monotonic or not  \cite{M2022}.
In systems with nonmonotonic covariances, described by the second line of Eqs.~(\ref{Taveragemon}), the MET is exponentially longer [for $\kappa(\ell_*)<0$] or exponentially shorter  [for $\kappa(\ell_*)>0$] than the MET for MD covariances with the same variance, as described by the first line of Eqs.~(\ref{Taveragemon}).

An independent support for the predictions~(\ref{PdV}) comes from  the bivariate normal distribution of $V(x)$. The joint distribution of our potential $V(x)$ taking some values $V_1$ and $V_2$ at two spatial points, separated by distance $2L$, is given by \cite{Gnedenko}
\begin{widetext}
\begin{equation}\label{bivariate}
P(V_1,V_2)=\frac{1}{2\pi \sqrt{\kappa^2(0)-\kappa^2(2L)}}\exp\left[-\frac{\kappa(0)z}{2(\kappa^2(0)-\kappa^2(2L))}\right],
\end{equation}
\end{widetext}
where
\begin{equation}\label{z}
z=V_1^2+V_2^2-2\frac{\kappa(2L)}{\kappa(0)} V_1 V_2.
\end{equation}
In particular, for the configuration where $V_1=-\Delta V/2$ and $V_2=\Delta V/2$ \cite{without}, Eqs.~(\ref{bivariate}) and (\ref{z}) yield
\begin{equation}
P(\Delta V)=\frac{1}{2\pi \sqrt{\kappa^2(0)-\kappa^2(2L)}}\exp\left[-\frac{\Delta V^2}{4(\kappa(0)-\kappa(2L))}\right],
\label{bound}
\end{equation}
for arbitrary $\Delta V$ and $L$. Clearly, Eq.~(\ref{bound}) provides an upper bound on the $\Delta V \gg 1$ tail of $P(\Delta V)$: the tail that we are interested in. This is because this equation accounts for \emph{all} possible configurations of $V(x)$ obeying the conditions $V(x=-L)=-\Delta V/2$ and $V(x=L)=\Delta V/2$,  regardless of whether they meet the additional conditions (\ref{defminusL})-(\ref{monotonV}) or not. Still, and somewhat surprisingly, the exponential factor in Eq.~(\ref{bound}) perfectly coincides  with the OFM action (\ref{act}) (where one should set $E=0$), leading to Eq.~(\ref{PdV}) for the MD and NM covariances, respectively.

\subsection{Nonzero bias}
\label{Monotone}

\begin{figure}[ht]
\includegraphics[width=0.30\textwidth]{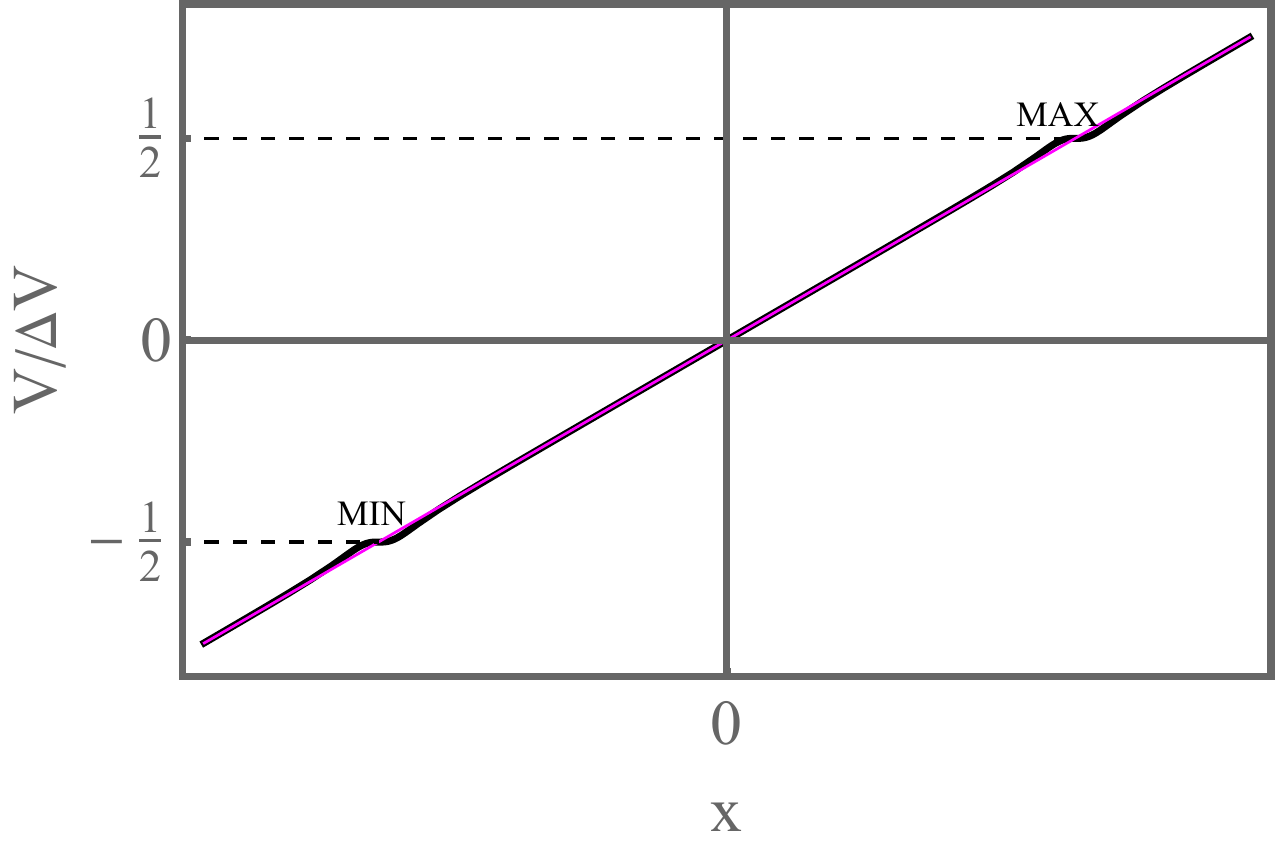}
\caption{Black line: an almost zero-action configuration of the disorder potential $V(x)$,
conditioned on the potential barrier $\Delta V$, for $E<0$ and $L=\Delta V/2|E|$. The straight magenta line shows
$V(x)=|E|x$. The potential stays arbitrarily close
to the straight line, except for infinitesimally small variations to accommodate the
extremum points.}
\label{graph_monotinic_and_positive}
\end{figure}

A nonzero bias breaks the left-right symmetry of the system. For concreteness, we continue to assume that the direction of escape is from left to right. There is a major difference between the
negative ($E<0$) and positive ($E>0$) bias.  For a negative bias the minimum of the action functional (\ref{first_action_with_bias}) can be made arbitrary small. To achieve this, the optimal configuration of the random component of the potential, $v(x)$, should stay very close to zero and, via infinitesimally small variations near $x=-L$ and $x=L$, create local minimum and a maximum, respectively. In its turn, $L$ has to be chosen to be close to $\Delta V/2|E|$ to provide the desired potential barrier $\Delta V$, see Fig. \ref{graph_monotinic_and_positive}. As a result, the probability of finding high barriers against the bias is quite high, and certainly
beyond the applicability of the OFM. Therefore, here we will only deal with a positive bias, which corresponds to the  particle escape along the bias.

\subsubsection{Monotonically decreasing covariance}
\label{Monotone1}

Let us first examine how the presence of a small bias affects the optimal value of $L=L(E)$ as described by Eq. (\ref{close_eq_1}).
One can see from Eq.~(\ref{close_eq_1}) that, as $E$ goes to zero, $\kappa'(2L)$ also goes to zero, so that the optimal barrier width $2L$ goes to infinity \cite{LV,M2022}. However, as one can check \textit{a posteriori}, it does so slower than $1/E$, that is $\lim_{E\rightarrow0^{+}}EL\left(E\right)=0$, and we will rely on this property.

When $E$ is sufficiently small, we can solve Eq.~(\ref{close_eq_1}) for $L$ perturbatively by using the $z\gg 1$ asymptotic of $\kappa'(z)$.
Also, for expected large $L$
we can neglect the term $\kappa\left(2L\right)$ compared with $\kappa\left(0\right)$. As a result, and in the leading order at small $E$, the optimal barrier width is
\begin{equation}
2L(E)\simeq \left(\left|\kappa'\right|\right)^{-1}\left[\frac{2E\kappa\left(0\right)}{\Delta V}\right]\,,\label{L_mon_neg}
\end{equation}
where $\left(\left|\kappa'\right|\right)^{-1}$ is the inverse function of the $z\gg 1$ asymptotic of $\left|\kappa'\right|$.

As a simple and useful example, let us suppose that the covariance decays as a power law,
$\kappa (z\to \infty)\simeq B\,z^{-\alpha}$, where $\alpha>0$.
Then Eq.~(\ref{L_mon_neg}) yields
\begin{equation}
    2L(E)\simeq \left[\frac{\alpha B \Delta V}{2 E \kappa(0)}\right]^{\frac{1}{\alpha+1}}\,.
    \label{L_power}
\end{equation}
As one can see, $\lim_{E\rightarrow0}EL\left(E\right)$ indeed vanishes, as we assumed.
In general, the faster the covariance goes to zero at large distances, the slower will $L$ tend to infinity when $E\to0$.

In the leading order in the bias $E$, the action (\ref{action_after_sol_1}) becomes
\begin{eqnarray}
   \!\! S(\Delta V) &\simeq& \frac{\left(\Delta V\right)^2}{4 \kappa(0)}\Biggl[ 1+\nonumber \\
    &+&\!\left(2+\frac{B}{\kappa(0)}\right)\! \left(\frac{\alpha  B}{2 \kappa(0)}\right)^{\frac{1}{\alpha +1}} \!\left(\frac{E}{\Delta V}\right)^{\frac{\alpha }{\alpha +1}}\Biggr].
    \label{act_power}
\end{eqnarray}
The nonanalytic correction $\sim E^{\frac{\alpha}{\alpha+1}}$, coming from the bias, describes
an increase in the ``action cost" of creating the barrier $\Delta V$ and, as a result, a decrease in the probability $\mathcal{P}(\Delta V)$ of observing this barrier. This perturbative calculation demands the strong inequality
$E\ll \Delta V$. When $E\rightarrow 0$,  the first line of Eq.~ (\ref{PdV}) is reproduced, as to be expected.

Now we can evaluate the MET from Eq.~(\ref{Tdoubleav1}). Substituting Eq.~ (\ref{act_power}) into Eq.~ (\ref{phi}),
we obtain
\begin{eqnarray}
&&\phi\left(\Delta V\right)=\frac{\Delta V}{D}-\frac{\left(\Delta V\right)^2}{4 \kappa(0)}\Biggl[1+\nonumber \\
    &&+\left(2+\frac{B}{\kappa(0)}\right) \left(\frac{\alpha  B}{2 \kappa(0)}\right)^{\frac{1}{\alpha +1}} \left(\frac{E}{\Delta V}\right)^{\frac{\alpha }{\alpha +1}}\Biggr]\,.\label{phi_Lorentzian}
\end{eqnarray}
We find the saddle point by minimizing this expression over $\Delta V$.
In the zeroth order in the bias we obtain $\Delta V=2 \kappa(0)/D$. We proceed perturbatively and, in the first order, obtain
\begin{equation}
  \Delta V^*\simeq  \frac{2 \kappa(0)}{D}-\frac{(\alpha +2) (B+2 \kappa(0)) }{2 (\alpha +1) \kappa(0)}\left(\frac{\alpha  B E^\alpha}{D}\right)^{\frac{1}{\alpha +1}}.
\end{equation}
As a result, we arrive at the following expression for the MET (\ref{Tdoubleav1}) in the presence of a positive bias:
\begin{equation}
\ln\left\langle T\right\rangle \simeq\frac{\kappa(0)}{D^2}-\frac{(\alpha  B)^{\frac{1}{\alpha +1}} (B+2 \kappa(0)) }{2 \kappa(0) D^2}(D E)^{\frac{\alpha }{\alpha +1}}\,,\label{time_for_Lorentzian}
\end{equation}
Crucially, as $D$ goes to zero, this MET is \emph{exponentially} smaller than its zero-bias counterpart.
Also noticeable are the nonanalytic scalings of the correction with the bias $E$ and with the diffusion coefficient $D$.
For $E\rightarrow0$, Eq.~ (\ref{Taveragemon}) is reproduced.

Repeating these calculations for a general MD covariance, we arrive at the following results for the action and the MET:
\begin{equation}
S\left(\Delta V\right) =\frac{1}{4}\frac{\left(\Delta V\right)^{2}+2E\Delta V\left(\left|\kappa'\right|\right)^{-1}\left[\frac{2E\kappa\left(0\right)}{\Delta V}\right]}{\kappa\left(0\right)-\kappa\left\{\left(\left|\kappa'\right|\right)^{-1}\left[\frac{2E\kappa\left(0\right)}{\Delta V}\right]\right\}}\,,
 \label{general_action_neg_m}
\end{equation}
\begin{equation}
\ln\left\langle T\right\rangle\simeq\frac{\kappa(0)-\kappa\left[\left(\left|\kappa'\right|\right)^{-1}
 \left(ED\right)\right]-DE\left(\left|\kappa'\right|\right)^{-1}\left(ED\right)}{D^{2}}.
 \label{general_time_neg_m}
\end{equation}
The saddle-point in this case is
\begin{align}
\Delta V_{*} =\frac{2\kappa\left(0\right)-2\kappa\left[\left(\left|\kappa'\right|\right)^{-1}\left(ED\right)\right]}{D}-
E\left(\left|\kappa'\right|\right)^{-1}\left(ED\right)\,.\label{dV_pos_E}
\end{align}

Another instructive example, and a consistency check, is provided by the Lorentzian covariance: $\kappa\left(z\right)=\left(1+ z^{2}/\sigma^2\right)^{-1}$,
where $\sigma$ is the characteristic correlation length. In this example $\kappa(0)=1$, $B=\sigma^2$ and $\alpha=2$. This example allows for an exact analytical solution of Eq. (\ref{close_eq_1}), valid for any value of $E>0$.  After some straightforward algebra we obtain:
\begin{equation}
2L = \left({\frac{\sigma^2\Delta V}{E}}\right)^{1/3}.\label{L_lor}
\end{equation}
Remarkably, this exact result perfectly coincides with the large-$L$ asymptotic~(\ref{L_power}) for this particular case.

\begin{figure}[ht]
\includegraphics[width=0.33\textwidth]{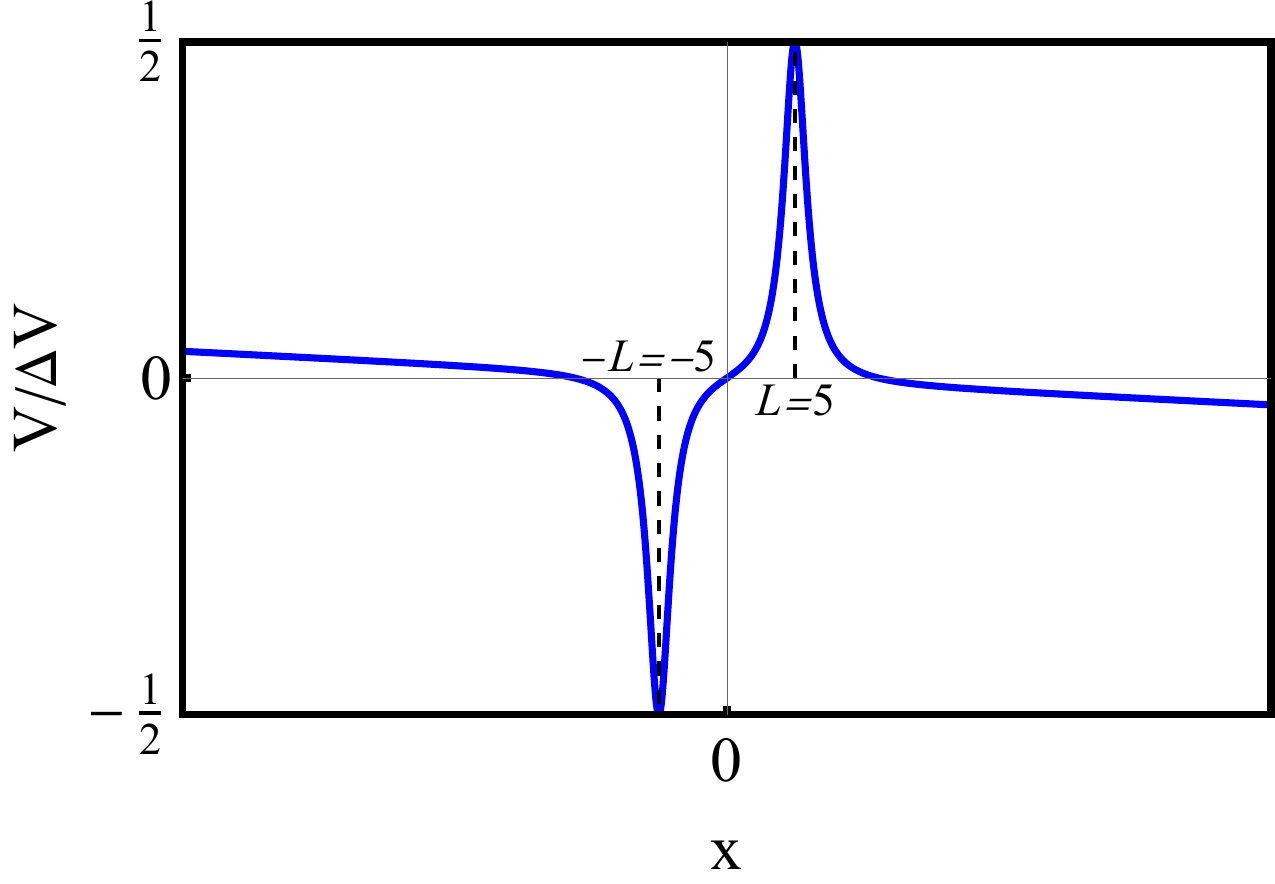}
\caption{Optimal configuration of the disorder potential $V(x)$
as described by Eq.~(\ref{L_lor}) for the Lorentzian covariance $\kappa\left(z\right)=\left(1+z^{2}/\sigma^2\right)^{-1}$, where, for simplicity, we set $\sigma=1$. For the chosen ratio $E/\Delta V =10^{-3}$ the optimal barrier width is $2L=10$.}
\label{graph_lorentsian_pos_bias}
\end{figure}

Going back to  Eqs.~(\ref{L_lor}) and~(\ref{action_after_sol_1}),
we obtain an exact expression for the action in this case:
\begin{equation}
S(\Delta V)=\frac{1}{4} \left[\left(\Delta V\right)^{2/3}+\left(\sigma E\right)^{2/3}\right]^3.\label{action_Lorentzian}
\end{equation}
The optimal disorder configuration (\ref{potential_after_sol_1}), with $L$  given by Eq.~(\ref{L_lor}), is shown in Fig. \ref{graph_lorentsian_pos_bias}. Visible is a finite barrier width $2L$. The localization of the barrier in this case which  should be contrasted with the delocalized barrier, $L\to \infty$, predicted for the zero bias.

The small-$E$ expansion of the action (\ref{action_Lorentzian}),
\begin{equation}
S(\Delta V) \simeq \frac{1}{4}(\Delta V)^2+\frac{3}{4}(\Delta V)^{4/3} \left(\sigma E\right)^{2/3},
\label{action_L_exp}
\end{equation}
perfectly agrees with the asymptotic presented in Eq.~(\ref{act_power}).

Let us summarize this subsection. Adding a positive bias to a disorder potential with monotonically decreasing covariance makes the action and, as a result, the MET sensitive not only
to the bias itself (as to be expected), but also to the large-distance behavior of the covariance derivative. The width of the optimal barrier configuration becomes finite, and it increases quite slowly as the bias goes to zero, see \textit{e.g.} Eq.~(\ref{L_power}). Finally, we predict nonanalytic scalings with the bias $E$ and with the diffusion coefficient $D$ of the (exponentially large) correction to the MET.

\subsubsection{Nonmonotonic covariance}
\label{NonMonotone}

For NM covariances $\kappa(z)$, a positive bias reduces the optimal value of $L$ as described by Eq.~(\ref{close_eq_1}).
When $E$ is small, Eq.~(\ref{close_eq_1}) in the leading order becomes
\begin{equation}
\kappa'\left(2L\right)\simeq -\frac{2E\left[\kappa\left(0\right)-\kappa\left(2L\right)\right]}{\Delta V}\,.\label{close_eq_2}
\end{equation}
We look for the optimal barrier width $2L = \ell_*+\delta \ell$ and solve Eq.~(\ref{close_eq_2}) perturbatively for the small correction $\delta \ell$.
In the leading order, we obtain
\begin{equation}
  2L=\ell_*-\frac{4 E}{\Delta V} \frac{\kappa(0)-\kappa(\ell_*)}{\kappa''(\ell_*)}.
  \label{dl}
\end{equation}
The small parameter of this perturbative expansion is $E\ell_*/\Delta V\ll 1$.
Equation~(\ref{dl}) shows that the positive bias ``squeezes" a bit the optimal disorder configuration $V(x)$.

Substituting Eq.~(\ref{dl}) into Eq.~(\ref{action_after_sol_1}) and expanding in the small parameter $E \ell_*/\Delta V$, we arrive at the following action
\begin{equation}
     S(\Delta V,E)\simeq\frac{\left(\Delta V\right)^2}{4 [\kappa(0)-\kappa(\ell_*)]}\left(1+2\frac{ E \ell_*}{\Delta V}+\ldots\right).
\label{act_nonm}
\end{equation}
Using Eqs.~(\ref{phi}) and (\ref{act_nonm}), we obtain the saddle point
\begin{equation}\label{saddle10}
\Delta V^*=\frac{2[\kappa\left(0\right)-\kappa\left(\ell_*\right)]}{D}-E \ell_*\,,
\end{equation}
which results in the MTE
\begin{equation}
    \ln\langle T\rangle \simeq \frac{\kappa(0) - \kappa(\ell_*)}{D^2} - \frac{E \ell_*}{D}.
\end{equation}
Contrary to the case of MD covariance, here the correction coming from the bias is analytic and not as prominent.

\section{Large deviation simulations}
\label{simulations}

\subsection{Simulation method}
\label{Details}

To generate numerical realizations of a one-dimensional statistically homogeneous Gaussian field (HGF) $V(x)$, we consider a discrete array  $\Vec{V}=(V(1), V(2),\ldots, V(M))^T$ of size $M\gg 1$, which provides a space discretization of the continuous field with the lattice step $\Delta x=1$. There is a straightforward method of sampling a discretized HGF $\Vec{V}$ with a given covariance matrix $C_{ij} = \kappa(|i-j|)=\langle V(i)V(j)\rangle$ for $i,j=\overline{1,M}$. The method consists of two steps: diagonalization of the covariance matrix $C_{ij}$ and a matrix multiplication:
\begin{equation}
    \vec{V}=C^{1/2}\vec{\xi}\,,
\label{NS1}
\end{equation}
where  $\Vec{\xi}=(\xi_1,\xi_2,\ldots,\xi_M)^T$ is a vector composed of independent standard normals. Although being transparent, this method is highly inefficient, since it involves two computationally expensive steps: the matrix diagonalization which requires $\mathcal{O}(M^3)$ operations, and matrix products which requires $\mathcal{O}(M^2)$ operations.

Instead of the poorly scalable matrix multiplications, a more efficient way to sample HGFs is to use the Circulant Embedding method (CEM) \cite{Chan1994,Dietrich1997}. The method is based on embedding a covariance matrix $C_{i,j}$ of size $M\times M$ in a larger circulant covariance matrix $C'_{ij}$ of size $M'>2M$. Then, using the fact that a matrix multiplication with a circulant matrix implements convolution,  one can replace the matrix equation (\ref{NS1}) with multiplication in Fourier space. Therefore the required data (like the result of matrix diagonalization) can be easily pre-computed numerically by using the fast Fourier transform (FFT), while sampling a HGF requires only $\mathcal{O}(M' \ln M')$ operations. In our implementation of the CEM we allowed for free ends of the sampled HGF $V(x)$. One could instead use periodic boundary conditions $V(x+M)=V(x)$ \cite{periodic}. See also Refs. \cite{Hartmann_2,Hartmann_1} for instructive implementations of the CEM for large deviation simulations of fractal Brownian motion (whose time derivative is a stationary Gaussian process).

Here we are interested in extremely small probability densities, which are virtually impossible to reach with conventional Monte Carlo (MC) simulations employing the Metropolis-Hastings algorithm. Therefore, in order to reach the large-$\Delta V$ tail of the distribution $\mathcal{P}(\Delta V)$, we employed the Wang-Landau (WL) algorithm \cite{WL1, WL2}.  Unlike the ordinary Metropolis-Hastings algorithm, where the acceptance/rejection decisions are Markovian, the WL algorithm takes into account information about previously visited states in such a way that it ``forces" the algorithm to explore the available configurational space more quickly.

In a nutshell, the WL algorithm aims at estimating the density of states $\mathcal{P}(\Delta V)=\exp[-S(\Delta V)]$ on a chosen interval $a\leq\Delta V\leq b$, updating on each MC step the histogram of visited states $H(\Delta V)$  and adjusting the action $S(\Delta V)$ in an iterative manner, see Ref. \cite{Landau_Binder} for details.

At the start of the simulation, the histogram is initialized to zero, $H(\Delta V)=0$, and the action is set to some guess function (we use $S(\Delta V)=1$). Let $\bigl\{\Vec{\xi}_{\text{r}},\Vec{V}_{\text{r}},\Delta V_{\text{r}}\bigl\}$  represent the running configurations of the random vector, the disorder potential computed using the CEM, and the maximal potential barrier of $\Vec{V}_{\text{r}}$, respectively. A proposed configuration of the random vector $\Vec{\xi}_{\text{p}}$ is generated by changing a randomly chosen component $[\Vec{\xi}_{\text{r}}]_j$ of the random vector according to the  Gaussian distribution centered at $[\Vec{\xi}_{\text{r}}]_j$:
\begin{equation}
    [\Vec{\xi}_{\text{p}}]_j\sim \exp[-(x-[\Vec{\xi}_{\text{r}}]_j)^2/2].
\end{equation}
Then, the proposed configuration of the disorder potential $\Vec{V}_{\text{r}}$ and its maximal potential barrier $\Delta V_{\text{r}}$ are computed using the CEM and the proposed random vector $\Vec{\xi}_{\text{p}}$.

Every decision on whether to accept (a) or reject (r) the proposed configuration $\bigl\{\Vec{\xi}_{\text{p}},\Vec{V}_{\text{p}},\Delta V_{\text{p}}\bigl\}$ is made according to the transition probability
\begin{eqnarray}
& p_{\text{a/r}}=\min\left(r(\Vec{V}_{\text{p}}\vert \Vec{V}_{\text{r}})\frac{\exp[-S(\Delta V_{\text{r}})]}{\exp[-S(\Delta V_{\text{p}})]},1\right),\\
 & \text{where }  r(\Vec{V}_{\text{p}}\vert \Vec{V}_{\text{r}})=\exp\left[\frac{\left[\Vec{\xi}_{\text{r}}]_j^2-
 [\Vec{\xi}_{\text{p}}\right]_j^2}{2}\right]\,.\nonumber
\end{eqnarray}
If the proposed configuration is rejected, the running configuration  $\bigl\{\Vec{\xi}_{\text{r}},\Vec{V}_{\text{r}},\Delta V_{\text{r}}\bigl\}$ is kept. Following each decision, the action and the histogram are updated
\begin{eqnarray}
 &  & S(\Delta V_{\text{r}})\to S(\Delta V_{\text{r}})-f\,,\\
 &  & H(\Delta V_{\text{r}})\to H(\Delta V_{\text{r}})+1\,,\nonumber
\end{eqnarray}
where $f$ is a modification factor (initially, we set $f=1$). This process is repeated until the histogram of visited states $H(\Delta V)$ is sufficiently flat. As a measure of flatness,  we used the condition
\begin{equation}
    0.9 \times \overline{H(\Delta V)} \leq \min (H(\Delta V)),
\end{equation}
where $\overline{H(\Delta V)}$ is the mean value of the histogram \textcolor{blue}{\cite{flat_note}}. Once this condition is met, $f$ is reduced, $f\to f/2$, and the histogram is reset to zero. Then, the process continues
until $f$ is sufficiently small.

It is known that, at early stages of a simulation,  the WL algorithm violates the detailed balance condition. The detailed balance is recovered, however, as the modification factor $f$ tends to zero \cite{Landau_Binder}. We stopped the simulations after 18 reductions of the modification factor $f$. The resulting relative accuracy of the simulations, $1-S^{(17)}(\Delta V)/S^{(18)}(\Delta V)$ was of the order of $10^{-3}$, where $S^{(i)}(\Delta V)$ is the numerical estimate of the action corresponding to the $i$-th reduction of the modification factor $f$. The remaining statistical errors of the WL algorithm, which are known to saturate at a nonzero value (which depends on the protocol of reduction $f$  \cite{Zhou, Belardinelli}), were of order $\mathcal{O}(1)$. In its turn, the numerically estimated action $S(\Delta V)$ varied from hundreds to thousands. The resulting relative accuracy was sufficient for our purposes.

\subsection{Simulation results}
\label{Results}

To verify our theoretical predictions for the tail of the barrier height distribution and the corresponding optimal configurations of the disorder for the monotonic and nonmonotonic covariances, we implemented the WL algorithm of sampling discretized configurations of the disorder Gaussian potential $V(x)$ on a regular lattice 
of length $M=10^3$  with the following three covariances:
\begin{equation}
\kappa(z)=
    \begin{cases}
         \frac{1}{1+\left(z/\sigma\right)^2}, \quad \text{MD,}\\
         \frac{3+2\cos(\omega z)}{5[1+\left(z/\sigma\right)^2]}, \quad \text{NMP,}\\
         \frac{\cos(\omega z)}{1+\left(z/\sigma\right)^2}, \quad \text{NMN.}
    \end{cases}
\label{Cov_func}
\end{equation}
\begin{figure}[ht]
\includegraphics[clip,width=0.35\textwidth]{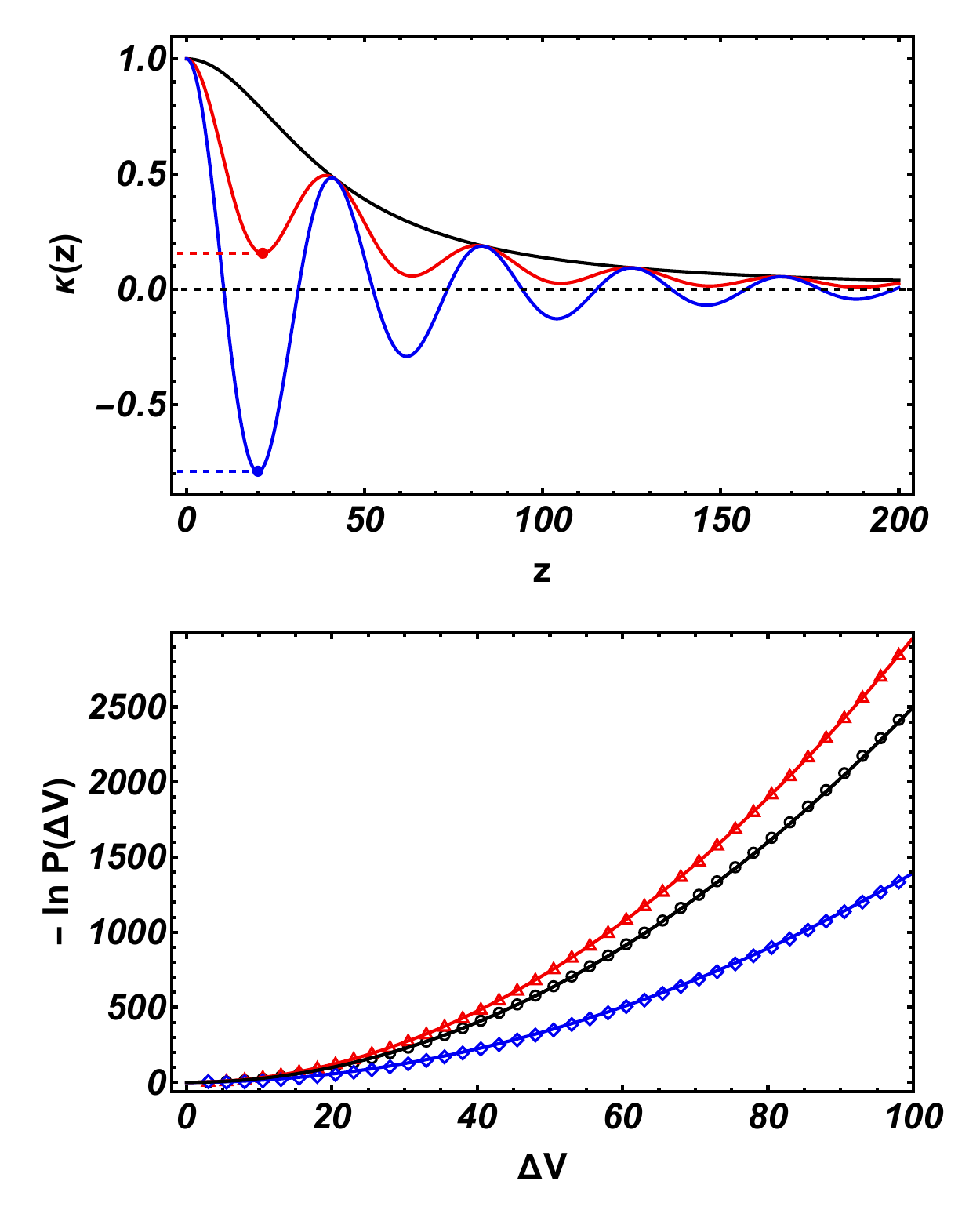}
\caption{Top panel: plotted versus the distance are three covariances, presented in Eq.~(\ref{Cov_func}): the MD covariance (black), the NMP covariance (red), and the NMN covariance (blue). The parameters of the first local minima for the red and blue curves are $\ell_*=21.41$ and $\kappa(\ell_*)=0.16$, and $\ell_*=20.06$ and $\kappa(\ell_*)=-0.79$, respectively. Bottom panel: The large-$\Delta V$ behavior of the rate function, $-\ln \mathcal{P}(\Delta V)$, as measured in the WL simulations of HGFs  with the MD covariance (black circles), the NMP covariance (red triangles), and  the NMN covariance (blue squares) for zero bias.  The theoretical predictions~(\ref{PdV}) are shown by the solid curves of the corresponding color.}
\label{zero_bias_P}
\end{figure}

These covariances are depicted in the top panel of Fig. \ref{zero_bias_P}. Notice that, for all the covariances (\ref{Cov_func}), the corresponding variances  are equal to $1$. The parameters $\sigma=40$ and $\omega=0.15$ in Eq.~(\ref{Cov_func}) represent  the correlation length and oscillation frequency, respectively, of this HGF. These values were chosen to be sufficiently large to accurately approximate a continuous HGF, but not too large so that effects of the finite system size would come into play.

\subsubsection{Zero bias}

We started with verifying the theory predictions for the unbiased potential \cite{LV,M2022}, which were summarized in Sec. \ref{zerobias}. The simulation results for the rate functions,  $-\ln \mathcal{P}(\Delta V)$, are shown in the bottom panel of Fig. \ref{zero_bias_P} alongside with the theoretically predicted rate functions given by Eq. (\ref{PdV}). As one can see, the agreement is excellent.

\begin{figure}[ht]
\includegraphics[clip,width=0.35\textwidth]{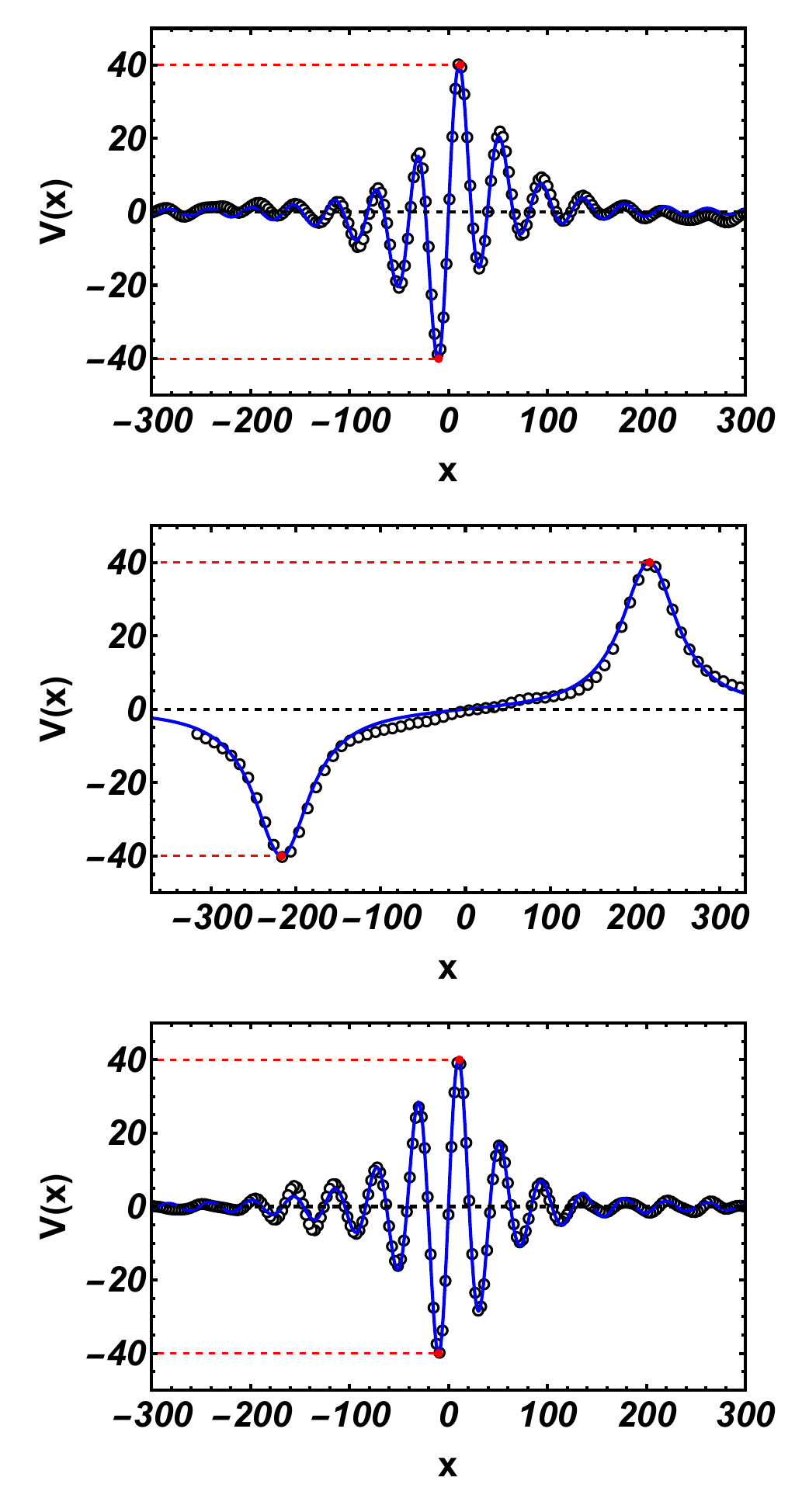}
\caption{Configurations of the disorder potential $V(x)$, corresponding to the potential barrier $\Delta V=80$, for three types of covariances (\ref{Cov_func}): simulations (black circles) vs. theoretical prediction~(\ref{V2}) (blue solid curve). Top: NMP covariance. Middle: MD covariance. Bottom: NMN covariance.}
\label{zero_bias_V}
\end{figure}

Examples of configurations of disorder potential $V(x)$, conditioned on large $\Delta V$ and sampled in the WL simulations \cite{shifts}, are presented in Fig. \ref{zero_bias_V}. Also shown are the optimal configurations predicted by Eq.~(\ref{V2}) with the optimal values of $L$. As one can see, the agreement of theory and simulations is excellent in all the three cases.  Notice that, for the NMN and NMP covariances, the theoretically predicted optimal barrier size is finite, allowing the true minimum of the action to be reached in the simulations. In the case of MD covariance, the true minimum of the action can be reached only when $L\to \infty$, and it is therefore inaccessible in numerical simulations. A finite value of  $L$, observed in the simulations, is caused by finite-size effects as it is comparable with the size of the simulated system.

\begin{figure}[ht]
\centering
\includegraphics[clip,width=0.35\textwidth]{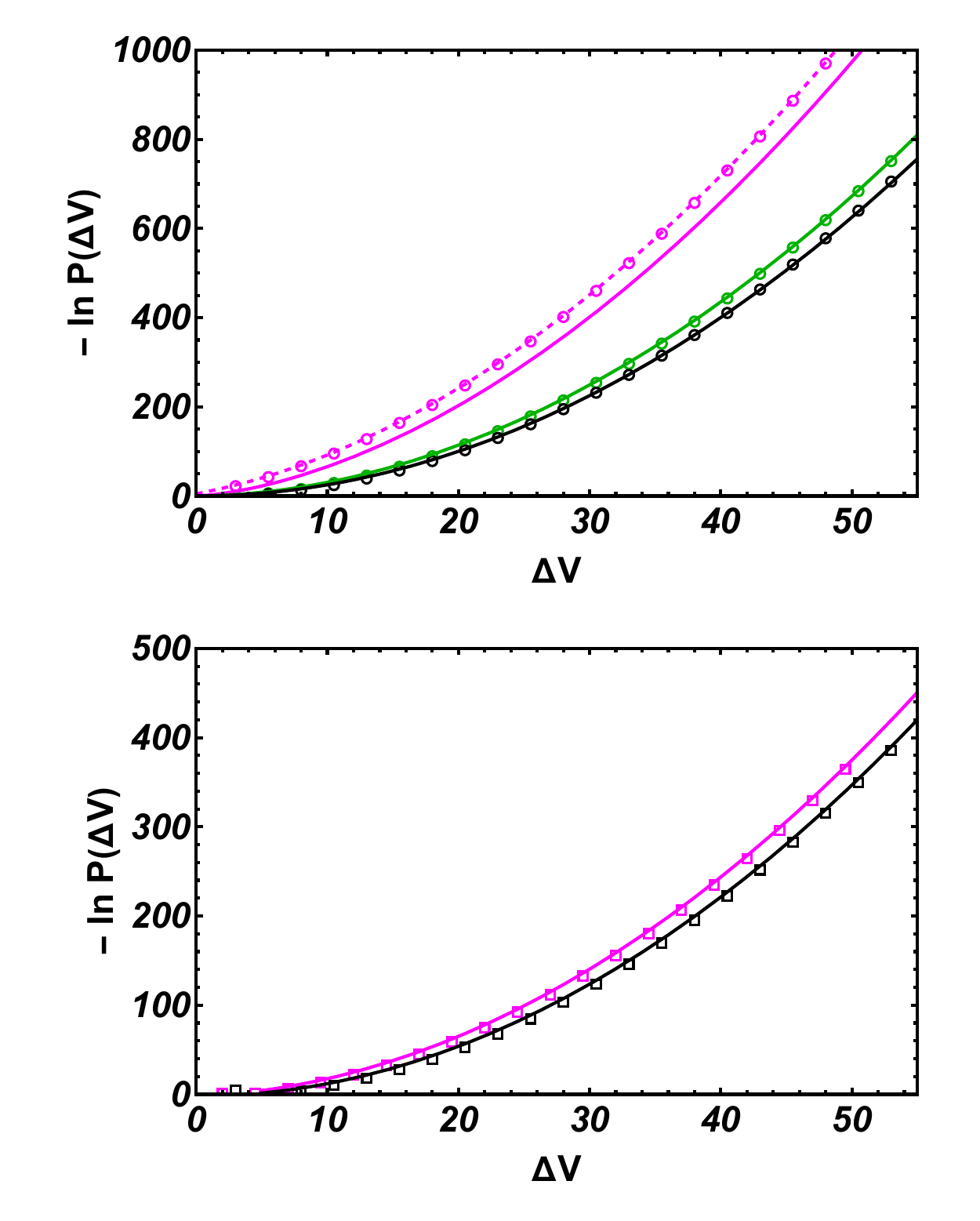}
\caption{Simulation results for the large-$\Delta V$ behavior of the rate function, $-\ln P(\Delta V)$, in the presence of a positive bias. Top panel: results for MD covariance (\ref{Cov_func}) for $E=0.005$ (green circles) and $E=0.1$  (magenta circles). For comparison, the results for the zero bias, $E=0$, are shown by black circles.  The theoretical predictions~(\ref{action_L_exp}) are shown by the solid curves of the corresponding color. The dashed magenta curve depicts the exact expression~(\ref{action_Lorentzian}). 
Bottom panel: results for the NMN covariance and $E=0.1$ (magenta squares). For comparison, the zero-bias results are shown by black squares. The solid curves of the corresponding color depict the theoretical prediction~(\ref{act_nonm}).}
\label{Positive_bias_P_rate}
\end{figure}

\subsubsection{Nonzero bias}

Now we present the results of a comparison of theory and simulations for the positively-biased potentials. The theoretical results were obtained in Secs. \ref{Monotone1} and \ref{NonMonotone}. The simulation results  for $-\ln \mathcal{P}(\Delta V)$ for the MD and NMN covariances, alongside with the theoretical predictions, are depicted in Fig. \ref{Positive_bias_P_rate}. Again, an excellent agreement is observed.

\begin{figure}[ht]
\centering
\includegraphics[clip,width=0.35\textwidth]{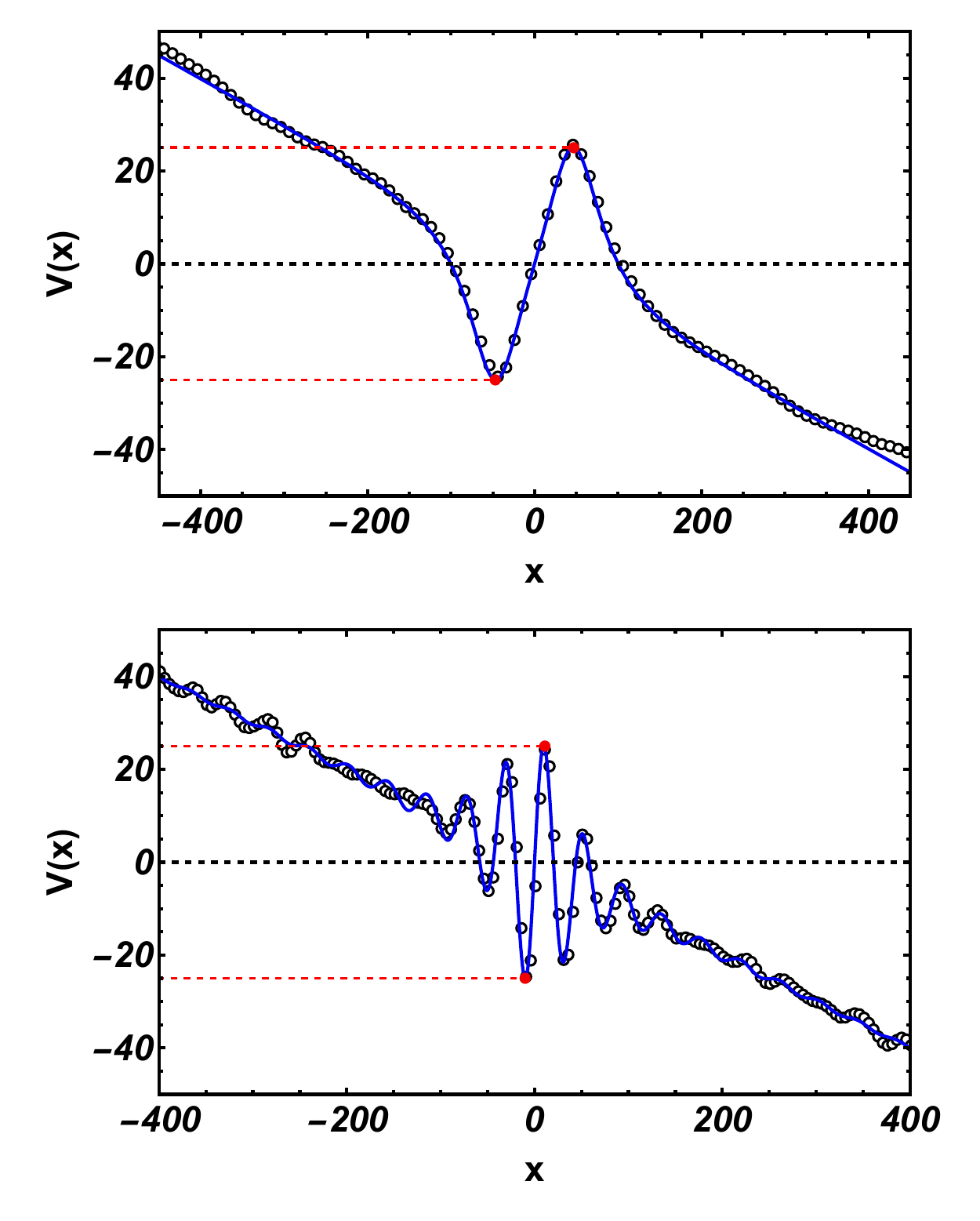}
\caption{Configurations of the disorder potential $V(x)$ with the bias $E=0.1$ and the potential barrier $\Delta V=50$ for the MD and NMN covariance (\ref{Cov_func}): simulations (black circles) vs. theoretical prediction  Eq.~(\ref{potential_after_sol_1}) (blue solid curve). The top panel
corresponds to the MD covariance, where the predicted $2L=\left(\sigma^2 \Delta V/E\right)^{1/3}\simeq 92.9$. The bottom panel corresponds to the NMN covariance, where the optimal value of $2L$ is given by Eq. (\ref{dl}). }
\label{Positive_bias_P_conf}
\end{figure}

Some examples of sampled configurations of the disorder potential $V(x)$ are presented in Fig. \ref{Positive_bias_P_conf}. In contrast to the unbiased case, the optimal distances between the \textit{spike} and \textit{antispike} are always finite here. Therefore, choosing a sufficiently large system size makes it possible to achieve the true minimum of the action in numerical simulations. In particular, one can clearly see the dramatic
effect of the bias (even a relatively small one, $E=0.1$) on the optimal barrier width $2L$ for the MD covariance, in a very good agreement with Eq.~(\ref{L_lor}). For comparison, the same bias $E=0.1$ hardly changes the optimal barrier width for the NM covariances.

\section{\label{SD}Summary and Discussion}

We found that the presence of a small potential bias leads to  an exponentially large reduction in the MET of overdamped particles trapped in local potential minima. The leading-order correction, which describes this reduction, behaves differently in disorder potentials with monotonic and nonmonotonic covariance.

In the nonmonotonic case, the effect of the bias can be accounted for via a perturbative expansion in the bias. In the monotonic case, the scaling of the MET with the bias is nontrivial, as it is affected by the large-distance asymptotic of the inverse function of the derivative of the covariance. The optimal barrier width of the biased potential in this case is finite, in contrast to the unbiased case, where it is infinite. Even a very small potential bias has a strong effect on the characteristic barrier width. As a result, all bias-related effects are more pronounced for disorder potentials with monotonically decreasing covariances.

We verified in numerical simulations our predictions for  the large-$\Delta V$ tail of the barrier height distribution, as well as earlier predictions  of this tail for zero bias  \cite{LV, M2022}. The simulations employed the WL algorithm and the circulant embedding method of sampling a homogeneous Gaussian field. We measured the large-$\Delta V$ tail of the barrier distribution $\mathcal{P}\left(\Delta V\right)$ for different covariances and bias magnitudes. The method also allowed us to sample the disorder potentials $V(x)$, allowing for a direct comparison with the OFM predictions for the optimal configurations, demonstrating excellent agreement.  Combining the WL algorithm with the circulant embedding method, we were able to measure probability densities below $10^{-1200}$.
The numerical methods, which we employed here, should be also useful when studying large deviation statistics of other Gaussian processes and fields.

Among future directions is an extension of theory to higher spatial dimensions, where the character of activated escape changes considerably. Indeed, in this case the particle must reach the closest saddle point of the random potential, rather than the closest maximum.

Finally, it is worth reminding that our results for the distribution tail of $\mathcal{P}(\Delta V)$ are readily applicable to the distribution tail of the differences between adjacent maxima and minima of correlated Gaussian processes in time.

\section*{Acknowledgments}

We are grateful to A. K. Hartmann and S. N. Majumdar for useful discussions. This research was supported by the Israel Science Foundation (Grant No. 1499/20).


\begin{thebibliography}{99}
\bibitem{aboutLV} In fact, Lopatin and Vinokur \cite{LV} evaluated the large-$T$ tail of the distribution of escape times $T$, but the latter distribution is simply related -- by Eq.~(\ref{Kramerstime}) -- to the distribution of the potential barriers $\Delta V$.
\bibitem{difference} We demand that the covariance $\kappa(z)$, and hence the inverse kernel (\ref{inverse}),  be a positive-definite function. This is a necessary and sufficient condition for the action (\ref{actiongeneral}) to be non-negative. In the nonmonotonic case we also demand that the difference $\kappa(0)-\kappa(\ell_*)$, where $\ell_*$ is the position of the closest-to-zero minimum of $\kappa(z)$, be greater than the differences between the rest of adjacent maxima and minima (if any) of the covariance.
\bibitem{arbitraryconstant} Due to translational symmetry with respect to $x$, there is in fact an infinite family of optimal configurations of the potential, conditioned on having a barrier $\Delta V$. That is, if $V_*(x)$ is a solution of the minimization problem, then $V_*(x+C)$,  for any real $C$, is also a solution. This fact, however, can only affect the pre-exponential factor of $\mathcal{P}(\Delta V)$ that we do not attempt to calculate here.
\bibitem{without} One can replace the condition $V_1=-V_2=\Delta/2$ by the more general condition
$V_2-V_1=\Delta V$ and then obtain the equality $V_1=-V_2=\Delta/2$ by minimizing the expression in the exponent of Eq.~(\ref{bivariate}) with respect to $V_1$.
\bibitem{flat_note}The histogram of visited states does not need to be perfectly flat. It usually suffices if the minimum of the histogram is about $0.8$ of the mean value \cite{Landau_Binder}. Choosing a higher value of this ratio would lower the statistical errors of the WL algorithm \cite{Belardinelli} at the expense of an increase of the computational time, which would be unnecessary in view of a good agreement with theory that we observed by setting the threshold at $0.9$.
\bibitem{shifts} In simulations with NM covariances we obtained a family of shifted optimal configurations $V_*(x+C)$ with different $C$, as to be expected from the translational symmetry \cite{arbitraryconstant}. The observed values of $C$ varied on the scale of the simulated system size $M$ (which was sufficiently large so that the effects of the boundaries were insignificant). 
    For the MD covariances the theoretically predicted minimum of the action is reached only in the limit of $L \to \infty$. In the simulated systems of finite size
    the observed optimal $L$ was comparable to the numerical system size. The latter was chosen to be sufficiently large to provide a good quantitative agreement with the theory. In both cases, we adjusted $C$ to zero for plotting purposes.
\end{thebibliography}
\end{document}